\documentclass[%
reprint, superscriptaddress,
amsmath,amssymb, aps,
]{revtex4-2}

\usepackage[T1]{fontenc} 

\usepackage[utf8]{inputenc}

\usepackage[english]{babel} \usepackage[babel]{microtype}

\usepackage{lipsum}
\usepackage{mathtools} 
 \usepackage{amssymb} \usepackage{amsfonts}
\usepackage{siunitx}  \usepackage{textgreek}

\usepackage[caption=false]{subfig}

\usepackage{xcolor} 

\definecolor{plotcol1}{RGB}{35,53,133}
\definecolor{plotcol2}{RGB}{232,104,5}
\definecolor{plotcol3}{HTML}{1b998b}
\definecolor{plotcol4}{HTML}{faffd8}
\definecolor{plotcol5}{HTML}{6a8d92}
\definecolor{plotcol6}{HTML}{4b4237}

\definecolor{divcol1}{HTML}{00429d}
\definecolor{divcol2}{HTML}{4771b2}
\definecolor{divcol3}{HTML}{73a2c6}
\definecolor{divcol4}{HTML}{ffbcaf}
\definecolor{divcol5}{HTML}{f4777f}
\definecolor{divcol6}{HTML}{cf3759}
\definecolor{divcol7}{HTML}{93003a}

\definecolor{shadecol2}{HTML}{a5d5d8}
\definecolor{shadecol1}{HTML}{a288a6}

\definecolor{supcol1}{HTML}{a6cee3}
\definecolor{supcol2}{HTML}{1f78b4}
\definecolor{supcol3}{HTML}{b2df8a}
\definecolor{supcol4}{HTML}{33a02c}
\definecolor{supcol5}{HTML}{fb9a99}
\definecolor{supcol6}{HTML}{e31a1c}
\definecolor{supcol7}{HTML}{fdbf6f}
\definecolor{supcol8}{HTML}{ff7f00}
\definecolor{supcol9}{HTML}{cab2d6}
\definecolor{supcol10}{HTML}{6a3d9a}

\definecolor{gold}{HTML}{FFD700}
\colorlet{silica}{blue!50}

\usepackage{acro} \acsetup{
  single=true 
} \DeclareAcronym{te}{
  short = TE,
  long = transverse electric,
  foreign-plural = {}
}
\DeclareAcronym{tm}{
  short = TM,
  long = transverse magnetic,
  foreign-plural = {}
}
\DeclareAcronym{tod}{
  short = TOD,
  long = third order dispersion,
  foreign-plural = {}
}
\DeclareAcronym{pcf}{
  short = PCF,
  long = photonic crystal fibre,
  foreign-plural = {}
}
\DeclareAcronym{sr-pcf}{
  short = SR-PCF,
  long = single-ring photonic crystal fibre,
  foreign-plural = {}
}
\DeclareAcronym{hcf}{
  short = HCF,
  long = hollow core fibre,
  foreign-plural = {}
}
\DeclareAcronym{wgm}{
  short = WGM,
  long = whispering gallery mode,
  foreign-plural = {}
}
\DeclareAcronym{nir}{
  short = NIR,
  long = near infrared,
  foreign-plural = {}
  }
\DeclareAcronym{cd}{
  short = CD,
  long = core diameter,
  foreign-plural = {}
}
\DeclareAcronym{fft}{
  short = FFT,
  long = Fourier transform,
  foreign-plural = {}
}
\DeclareAcronym{od}{
  short = OD,
  long = outer diameter,
  foreign-plural = {}
}
\DeclareAcronym{sem}{
  short = SEM,  
  long = scanning electron micrograph,
  foreign-plural = {}
}
\DeclareAcronym{rs}{
  short = RS,
  short-plural = {},
  long = rotation stage,
  foreign-plural = {}
}
\DeclareAcronym{ts}{
  short = TS,
  long = translation stage,
  foreign-plural = {}
}
\DeclareAcronym{mmf}{
  short = MMF,
  long = multimode fibre,
  foreign-plural = {}
}
\DeclareAcronym{zdw}{
  short = ZDW,
  long = zero dispersion wavelength,
  foreign-plural = {}
}
\DeclareAcronym{uv}{
  short = UV, 
  long = ultraviolet,
  foreign-plural = {}
}
\DeclareAcronym{euv}{
  short = EUV, 
  long = extreme ultraviolet,
  foreign-plural = {}
}
\DeclareAcronym{mir}{
  short = mid-IR,
  long = mid-infrared,
  foreign-plural = {}
}
\DeclareAcronym{gvd}{
  short = GVD,
  long = group velocity dispersion,
  foreign-plural = {}
}
\DeclareAcronym{spm}{
  short = SPM,
  long = self-phase modulation,
  foreign-plural = {}
}
\DeclareAcronym{hhg}{
  short = HHG,
  long = high harmonic generation,
  foreign-plural = {}
}
\DeclareAcronym{arr-pcf}{
  short = ARR-PCF,
  long = anti-resonant-reflection photonic crystal fibres,
  foreign-plural = {}
}
\DeclareAcronym{PDMS}{
  short = PDMS,
  long = polydimethylsiloxane,
  foreign-plural = {}
}
\DeclareAcronym{MoO}{
  short = \textalpha-MoO\textsubscript{3},
  long = \textalpha-molybdenum trioxide,
  foreign-plural = {}
}
\DeclareAcronym{VO}{
  short = \textalpha-V\textsubscript{2}O\textsubscript{5},
  long = \textalpha-vanadium oxide,
  foreign-plural = {}
}
\DeclareAcronym{AFM}{
  short = AFM,
  long = atomic force microscopy,
  foreign-plural = {}
}
\DeclareAcronym{FTIR}{
  short = FTIR,
  long = Fourier transform infrared,
  foreign-plural = {}
}
\DeclareAcronym{MCT}{
  short = MCT,
  long = mercury cadmium telluride,
  foreign-plural = {}
}
\DeclareAcronym{TMM}{
  short = TMM,
  long = transfer matrix method,
  foreign-plural = {}
}
\DeclareAcronym{FP}{
  short = FP,
  long = Fabry--Pérot,
  foreign-plural = {}
}
\DeclareAcronym{TO}{
  short = TO,
  long = transverse optical,
  foreign-plural = {}
}
\DeclareAcronym{LO}{
  short = LO,
  long = longitudinal optical,
  foreign-plural = {}
}
\DeclareAcronym{ER}{
  short = ER,
  long = extinction ratio,
  foreign-plural = {}
}
\DeclareAcronym{IL}{
  short = IL,
  long = insertion loss,
  foreign-plural = {}
}
\DeclareAcronym{vdW}{
  short = vdW,
  long = van der Waals,
  foreign-plural = {}
}

\usepackage{graphicx}
\usepackage{tikzexternal}
 \newcommand{\MoO}{\textalpha-MoO\textsubscript{3}}

\newcommand{\abs}[1]{\left|#1\right|} \newcommand{\nfrac}[2]{#1/#2}

\makeatletter \newcommand*{\rom}[1]{\expandafter\@slowromancap\romannumeral #1@}
\makeatother

\newcommand*{\figswitch}[1]{#1}

\usepackage{hyperref} \hypersetup{ hidelinks }

\addto\extrasenglish{%
   }

\begin{document}

\preprint{} \title{Deep-subwavelength Phase Retarders at Mid-Infrared
  Frequencies \\ with van der Waals Flakes}
\author{Michael T. Enders} \author{Mitradeep Sarkar}%
\author{Aleksandra Deeva} \author{Maxime Giteau} \author{Hanan Herzig Sheinfux}
\affiliation{ICFO --- Institut de Ciencies Fotoniques, The Barcelona Institute
  of Science and Technology, 08860 Castelldefels (Barcelona), Spain}
\author{Mehrdad Shokooh-Saremi} \affiliation{Department of Electrical
  Engineering, Ferdowsi University of Mashhad, Mashhad 91779-48944, Iran}
\author{Frank H.L. Koppens} \author{Georgia T. Papadakis}
\email{georgia.papadakis@icfo.eu} \affiliation{ICFO --- Institut de Ciencies
  Fotoniques, The Barcelona Institute of Science and Technology, 08860
  Castelldefels (Barcelona), Spain}

\date{\today}

\begin{abstract}
  Phase retardation is a cornerstone of modern optics, yet, at \ac{mir}
  frequencies, it remains a major challenge due to the scarcity of
  simultaneously transparent and birefringent crystals. Most materials
  resonantly absorb due to lattice vibrations occurring at \ac{mir} frequencies,
  and natural birefringence is weak, calling for hundreds of microns to
  millimeters-thick phase retarders for sufficient polarization rotation. We
  demonstrate \ac{mir} phase retardation with flakes of \ac{MoO} that are more
  than ten times thinner than the operational wavelength, achieving \num{90}
  degrees polarization rotation within one micrometer of material. We report
  conversion ratios above \SI{50}{\percent} in reflection and transmission mode,
  and wavelength tunability by several micrometers. Our results showcase that
  exfoliated flakes of low-dimensional crystals can serve as a platform for
  \ac{mir} miniaturized integrated polarization control.
\end{abstract}

\maketitle

\acresetall
\section{Introduction}

From understanding the early universe~\cite{kamionkowski2016araa} and mapping
our galaxy~\cite{Hu2023} to encoding quantum information in
photons~\cite{zhang2008pra}, the polarization of an electromagnetic field is an
important internal property of light. Controlling light's polarization is also
an indispensable part of modern life~\cite{Ding2021}, spanning applications in
sensing~\cite{Lee2023}, imaging~\cite{Zhou2022},
microscopy~\cite{Intaravanne2022}, and telecommunications~\cite{Wang2022}. Phase
retardation in particular, the process of introducing a phase difference between
two orthogonal polarization states, is a crucial building block of modern
optics. This phase difference is typically introduced by a phase retarder (or
wave plate), which converts between linear and circular polarization and can
achieve any elliptically polarized state in-between~\cite{osti_5951989}. Despite
the central role phase retardation plays in manipulating photons, for a large
portion of the \ac{mir} spectral range, namely at wavelengths above the central
wavelength of the CO\textsubscript{2} laser of \SI{10.6}{\micro\meter},
commercial wave plates are extremely scarce.

\begin{figure}[b]
  \flushleft \figswitch{\begin{tikzpicture}
  \node[shadecol1!80!black] at (1.6, 5.1) {\bfseries\footnotesize Deep sub-wavelength};
  \node[shadecol2!65!black] at (5.9, 5.1) {\bfseries\footnotesize Optically thick};
  \begin{axis}[       
    width = \linewidth,
    height = 0.75\linewidth,
    xlabel= {Phase retarder thickness, \(d\)}, 
    ylabel= {Wavelength, \(\lambda\)},
    x unit=\si{\micro\meter}, 
    y unit=\si{\micro\meter}, 
    xmin = 0.5, xmax = 5e2,
    ymin=0.22, ymax=45,
    xmode=log, ymode=log,
    log ticks with fixed point,
    ]
    \addplot+[domain = 0.1:100, samples=200, forget plot, draw=none, name
    path=subwl, opacity=0.5] {x};
    \draw[draw=none, name path=yAxisL] (axis cs: 0.1,0.1) -- (axis cs:
    0.1,40);
    \draw[draw=none, name path=yAxisR] (axis cs: 2000,0.1) -- (axis cs:
    2000,20);
    \addplot [fill=shadecol1, forget plot, opacity=0.5] fill
    between[of=subwl and yAxisL];
    \addplot [fill=shadecol2, forget plot, opacity=0.20] fill
    between[of=yAxisR and subwl];
    \addplot+[domain = 0.1:100, samples=200, draw=black, dashed] {x}
    node[above, black, pos=0.4, rotate=41] {\(d = \lambda\)};
    \pgfplotsset{cycle list shift=-1}
    \addplot+[black!50, very thick] table[col sep=comma]
    {data/fig1-material-comparison/MgF.csv} node[above,pos=0.55, rotate=35,
    yshift=-0.6mm] {
      MgF\textsubscript{2} 
    };
    \addplot+[black!50, very thick] table[col sep=comma]
    {data/fig1-material-comparison/SiO.csv} node[above,pos=0.85, rotate=46,
    yshift=-0.6mm] {
      SiO\textsubscript{2} 
    }; 
    \addplot+[black!50, very thick] table[col sep=comma]
    {data/fig1-material-comparison/CdSe.csv} node[above left,pos=0.8,
    rotate=42, yshift=-0.6mm] {
      CdSe\vphantom{g} 
    };
    \addplot+[black!50, very thick] table[col sep=comma]
    {data/fig1-material-comparison/AgGaSe.csv} node[above,pos=0.7,
    rotate=48, yshift=-0.7mm] {
      AgGaSe\textsubscript{2} 
    };
    
    \addplot+[black!50, very thick] table[col sep=comma]
    {data/fig1-material-comparison/AgGaS.csv} node[above, pos=0.8,
    rotate=48, yshift=-0.7mm] {
      AgGaS\textsubscript{2} 
    };
    \addplot+[black!50, very thick] table[col sep=comma]        
    {data/fig1-material-comparison/BaBO.csv} node[below,pos=0.75,
    rotate=44] {
      BaB\textsubscript{2}O\textsubscript{4} 
    };
    \addplot+[black!50, very thick] table[col sep=comma]        
    {data/fig1-material-comparison/Calcite.csv} node[above,pos=0.75,
    rotate=44, yshift=-0.6mm] {
      CaCO\textsubscript{3}
    };
    
    \addplot+[black!50, very thick] table[col sep=comma]        
    {data/fig1-material-comparison/BTS.csv} node[above,pos=0.55,
    rotate=41, yshift=-0.6mm] {
      BaTiS\textsubscript{3}
    };
    
    \addplot+[divcol7, very thick] table[col sep=comma]
    {data/fig1-material-comparison/MoO.csv} node[below,pos=0.85,
    rotate=5] {\MoO (\(x\))};
    \addplot+[divcol7, very thick] table[col sep=comma]
    {data/fig1-material-comparison/MoOy.csv} node[above,pos=0.6,
    rotate=10, yshift=-0.6mm] {\MoO (\(y\))};
  \end{axis}
\end{tikzpicture}}
  \caption{\label{fig1:material-comparison} \textbf{\Ac{mir} material
      comparison.} Thickness and operational wavelength range for half-wave
    plate operation in transmission for common and state-of-the-art anisotropic
    materials as compared to flakes of \ac{MoO}. The calculation is based on
    \autoref{eq:half-wave-plate}, and the optical constants for the considered
    materials are taken from: \cite{dodge1984aoa} for MgF\textsubscript{2},
    \cite{ghosh1999oc} for SiO\textsubscript{2}, \cite{lisitsa1969pssb} for
    CdSe, \cite{kato2021aoa} for AgGaSe\textsubscript{2}, \cite{kato1996jjap}
    for AgGaS\textsubscript{2}, \cite{tamosauskas2018omeo} for
    BaB\textsubscript{2}O\textsubscript{4}, \cite{ghosh1999oc} for
    CaCO\textsubscript{3}, \cite{niu2018np} for BaTiS\textsubscript{3}.}
\end{figure}
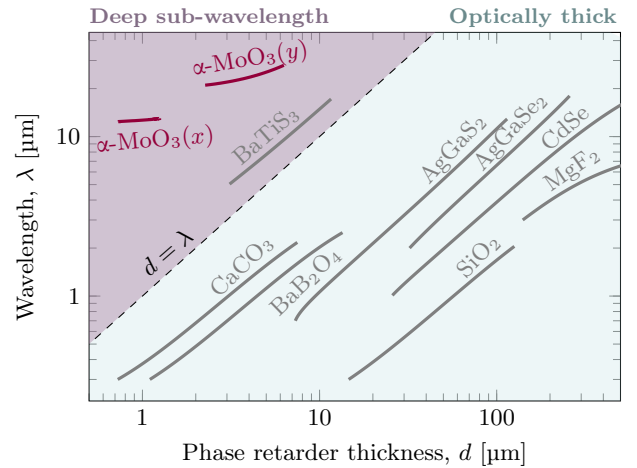

The lack of phase retardation schemes at mid-IR frequencies is an important
roadblock in photonics, as this spectral range is fundamental to both science
and technology, and currently experiencing rapid
advancements~\cite{Caldwell2015}. Plank’s law of thermal radiation predicts a
peak of thermal emission at wavelengths between \SIrange{10}{12}{\micro\meter}
for near-room temperatures, thus enabling a variety of light-harvesting
applications including contactless temperature regulation, renewable
energy~\cite{Datas2019,Papadakis2021,Xiao2022}, and thermal
camouflage~\cite{zhu2020lsa}. The \ac{mir} range is also crucial in astronomy,
for example \ac{mir} polarimetry enables probing hot interstellar matter in our
galaxy~\cite{Draine2021,Packham2008}. Additionally, the earth's atmosphere
becomes transparent in this range, making \ac{mir} night
vision~\cite{Konstantatos2018} and daytime radiative cooling~\cite{raman2014n}
feasible and relevant, as well as paving the way for \ac{mir}
communications~\cite{Zou2022}. Molecular vibrations also occur in the \ac{mir}
range, thereby \ac{mir} sensing~\cite{Bareza2022},
detection~\cite{Huo2017,Wang2023}, and medical microscopy~\cite{Kviatkovsky2023}
depend on controlling \ac{mir} photons.

Like molecular vibrations, however, crystal lattice vibrations (phonons) in
solids also occur at the mid-IR frequencies, yielding strong resonant absorption
in all polar dielectrics. This greatly limits the range of available materials
for polarization control elements and explains the technological gap in mid-IR
phase retarders. In particular, for efficient phase retardation, transparent
materials with strong intrinsic material birefringence, defined as
\(\Delta n = \abs{n_{\text{o}}- n_{\text{e}}}\), where \(n_{\text{o}}\) and
\(n_{\text{e}}\) are the ordinary and extraordinary refractive indices of a
uniaxial material, are required. Most materials that retain some degree of
birefringence at \ac{mir} frequencies resonantly absorb. Furthermore,
birefringence is a weak effect and remains below unity (\(\Delta n \ll 1\)) in
bulk crystals, introducing major scalability challenges in integrated \ac{mir}
photonics. To understand this, let us consider the principle of operation of a
phase retarder composed of a lossless uniaxial crystal. Its optimal thickness is
given by:
\begin{equation}
  \label{eq:half-wave-plate}
  d_{\lambda/\rho} = \frac{\lambda}{\rho\Delta n},
\end{equation}
where \(\lambda\) is the wavelength of operation. \autoref{eq:half-wave-plate}
gives the ideal thickness (\(d_{\lambda/\rho}\)) for half-wave plate ($\rho=2$)
or quarter-wave plate ($\rho=4$) operation, rotating linear polarization and
converting it to circular polarization, respectively. To compensate for weak
natural birefringence at \ac{mir} frequencies, wave plates with thicknesses
ranging from hundreds of microns to millimeters are required for sufficient
phase retardation~\cite{Dai2022}. This is shown in
\autoref{fig1:material-comparison}, which presents a map of available
birefringent materials that can operate as half-wave plates and the relevant
spectral range within which they remain transparent.

\begin{figure}[t]
  \flushleft\figswitch{\input{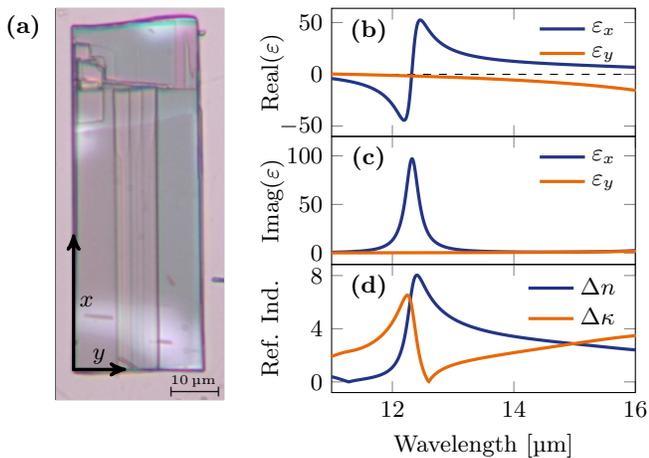}}
  \caption{\label{fig1:permittivity} \textbf{Molybdenum trioxide.}
    \textbf{(a)}~Microscope image of a \acs{MoO} flake where the scale bar is
    \SI{10}{\micro\meter}. The thickness of the flake ranges from
    \SI{0.65}{\micro\meter} on the right side to \SI{0.9}{\micro\meter} on the
    left. The \(x\)-direction represents the crystallographic direction
    \([100]\) and the \(y\)-direction represents
    \([001]\). \textbf{(b)}~Measured real and \textbf{(c)}~imaginary part of the
    permittivities of the in-plane (\(x-y\)) directions of \ac{MoO}. The method
    of extraction of these permittivities is described
    in~\cite{sarkar2023}. \textbf{(d)}~Birefringence
    \(\Delta n = \abs{n_{x} - n_{y}}\) and dichroism
    \(\Delta\kappa = \abs{\kappa_{x} - \kappa_{y}}\) of \acs{MoO}.}
\end{figure}

Conventional materials like CaCO\textsubscript{3} in the near-IR or
AgGaSe\textsubscript{2} in the mid-IR, shown in
\autoref{fig1:material-comparison}, ought to be hundreds of wavelengths thick to
perform as half-wave plates and fall within the class of ``optically thick''
components (cyan highlighted region). The only available substitute for mid-IR
wave plates are Fresnel rhomb retarders. These are even bulkier, approaching in
size the centimeter-scale. To reduce size, one can engineer strong birefringence
(\(\Delta n>1\)) for selected wavelengths using form birefringence, for example
in liquid crystals~\cite{Marco2021, herman2017lca}, or by design, using
topology-optimized metasurfaces~\cite{Shi2023}. Nonetheless, these methods often
require complex lithographic steps, and remain limited in their
scalability. Reducing the dimensions of \ac{mir} phase retardation schemes will
resolve major practical challenges including mechanical instabilities, beam
incoherence, overheating, thus enabling miniaturization for on-chip and
integrated optics applications, where device compactness is
required~\cite{Lin2022,Zhou2019}.

In this article, instead of conventional phase retardation schemes with lossless
anisotropic crystals, we consider a different class of materials as a platform
for mid-IR phase retardation: \ac{vdW}, low-dimensional, in-plane anisotropic,
highly dispersive materials. Several low-dimensional materials, such as
\ac{MoO}~\cite{ma2018n, zheng2018am,alvarez-perez2020am} and
\ac{VO}~\cite{taboada-gutierrez2020nm}, have been recently considered for their
in-plane hyperbolic dispersion, leading to a plethora of intriguing
phenomena. Examples include negative refraction~\cite{sternbach2023s, hu2023s}
and reflection~\cite{alvarez-perez2022sa}, reversed Cherenkov
radiation~\cite{guo2023nc}, and topological~\cite{hu2020n, duan2021sa},
configurable and tunable~\cite{chen2020nm, hu2023s}, steerable~\cite{dai2020nca}
phonon polaritons. By contrast, here, we utilize the hyperbolic response of
\ac{MoO} as an extreme case of birefringence to demonstrate a practical
mechanism: phase retardation, for which a hyperbolic response is beneficial but
not a requirement.

In particular, we measure experimentally the colossal resonant birefringence,
approaching \(\Delta n=8\) (\autoref{fig1:permittivity}), owing to the
ultra-long lifetimes of phonon polaritons in \ac{MoO} along orthogonal
crystallographic directions. This huge $\Delta n$ is ideal for polarization
control, as Folland and Caldwell commented in 2018~\cite{folland2018n}. By
\autoref{eq:half-wave-plate}, in the lossless-limit, increasing birefringence
\(\Delta n\) can reduce the thickness of a phase retarder. Considering the
measured birefringence of \ac{MoO} and its frequency dispersion, we show in
\autoref{fig1:material-comparison} that it is placed near the top-left corner of
the map (purple highlighted region), in the extreme sub-wavelength regime, where
the notation ``\(x\)'' and ``\(y\)'' corresponds to a phonon resonance along the
\(x\)- and \(y\)-crystal direction, respectively. This map shows that by
leveraging the birefringence of \ac{MoO}, the required thickness for sufficient
phase retardation reduces by up to three orders of magnitude with respect to the
operation wavelength, as compared to conventional bulk crystals. Recently,
thin-film polarizers composed of \ac{MoO} were demonstrated that reflect one
linear polarization and dissipate the other~\cite{sahoo2022aom}. Dereshgi
\textit{et al.} experimentally demonstrated a \ac{mir} phase retarder employing
\ac{MoO}~\cite{dereshgi2020nc}; however these results pertain to
heterostructure, requiring multiple steps of deposition and transfer. These
devices, only working in reflection, exhibited high resonant absorption due to
the optical phonons, and could only reflect a small portion (\SI{15}{\percent})
of the incoming light into the cross-polarized state.

In contrast, we probe experimentally the polarization rotation as well as the
ellipticity of mid-IR photons, and showcase that single \ac{vdW} flakes of
anisotropic crystals suffice and can serve a practical material platform for
miniaturized \ac{mir} phase retardation in either reflection or transmission,
with improved performance metrics. Despite the strong frequency dispersion and
resonant absorption of \ac{MoO} (\autoref{fig1:permittivity}c) near the
wavelength range of operation, insertion losses remain small due to the
negligible thickness of the flakes as compared to the wavelength. With respect
to the canvas of available materials, as a final remark, we note that a class of
quasi one-dimensional chalcogenides has recently been
introduced~\cite{niu2018np, mei2023}, having birefringence comparable to
\ac{MoO} in the \ac{mir}, as shown in \autoref{fig1:material-comparison} for
BaTiS\textsubscript{3}. Nonetheless, no phase retardation devices operating in
the \ac{mir} have been demonstrated so far, even with such material systems.

To our knowledge, our results mark the first reported phase retarders in the
mid-IR above \SI{11}{\micro\meter}. Our results also demonstrate the thinnest
reported phase retarders relative to the operation wavelength, for all spectral
ranges. Finally, these do not require any lithography, thus reducing
simultaneously design complexity and material volume with respect to existing
phase retardation schemes. The concepts shown in this work with \ac{MoO} for the
sake of demonstration also apply to \ac{VO} (see previous numerical results
in~\cite{dixit2021sr}), which has the same type of anisotropy as \ac{MoO} at
larger wavelengths~\cite{taboada-gutierrez2020nm}.

\begin{figure*}[]
  \flushleft \subfloat{\label{fig2:setup}
    \figswitch{\begin{tikzpicture}
  \fill[plotcol4!85!plotcol2, opacity=0.7] (-4, -0.3) rectangle (-1.7, 0.3);
  \fill[plotcol4!85!plotcol2, opacity=0.7] (-1, -0.3) -- (-0.45, 0) -- (-1, 0.3) -- cycle;
  \fill[plotcol4!85!plotcol2, opacity=0.7] (0.9, -0.3) -- (0.35, 0) -- (0.9, 0.3) -- cycle;
  \fill[plotcol4!85!plotcol2, opacity=0.7] (1.65, -0.3) rectangle (3.9, 0.3);  
  
  \draw[->, thick, plotcol1, opacity=0.8] (-4,0) node[left] {IR} -- (-1.75,0);
  \draw[->, thick, plotcol1, opacity=0.8] (-1,0) -- (-0.45,0);
  \draw[->, thick, plotcol1, opacity=0.8] (0.35,0) -- (0.9,0);
  \draw[->, thick, plotcol1, opacity=0.8] (1.65,0) -- (3.85,0);
  
  \fill[shadecol1] (0,-0.7) rectangle (-0.45,0.7) node[midway, black, rotate=90] {Silicon};
  \fill[shadecol2] (0,-0.7) rectangle (0.35,0.7) node[midway, black, rotate=90] {\acs{MoO}};
  \draw[<->] (0, -0.8) -- (0.35, -0.8) node[midway, below] {\(d\)};
  
  \fill[gray] (-1.5, 0.5) rectangle (-1, -0.5);
  \fill[gray!50!white] (-1.45, 0.5) rectangle (-1.3, -0.5);
  \fill[gray] (-1.75, -0.6) rectangle (-1.45, 0.6) node[above, xshift=1mm] {Condenser};
  \fill[gray!50!white] (-1.75, -0.6) rectangle (-1.7, 0.6);

  \fill[gray] (0.9, 0.5) rectangle (1.4, -0.5);
  \fill[gray!50!white] (1.35, 0.5) rectangle (1.2, -0.5);
  \fill[gray] (1.65, -0.6) rectangle (1.35, 0.6) node[above, xshift=-1mm] {Objective};
  \fill[gray!50!white] (1.6, -0.6) rectangle (1.65, 0.6);

  \fill[plotcol5] (3.85, -0.4) rectangle (4.55,0.4) node[above, xshift=-3mm] {Detector};

  \fill[plotcol3, opacity=0.8] (-2.9, -0.4) rectangle (-2.8, 0.4) node[above]
  {Pol.};
  \draw[thick, plotcol3, ->] (-3.05,0) [partial ellipse=-170:170:0.07cm and 0.4cm] node[pos=0.2, below] {\(\varphi_{1}\)};

  \begin{scope}[xshift=3mm]
  \fill[plotcol3, opacity=0.8] (2.6, -0.4) rectangle (2.7, 0.4) node[above] {Pol.};
  \draw[thick, plotcol3, ->] (2.45,0) [partial ellipse=-170:170:0.07cm and
  0.4cm] node[pos=0.2, below, ] {\(\varphi_{2}\)};
  \end{scope}

  \draw[dashed] (-2.25,0) -- (-2.25,0.9);
  \draw[dashed, ->] (-2.25,0.9) -- (-5.8,0.9);
  \begin{scope}[shift={(-6.5,0.5)}, scale=0.75]
    \begin{axis}[
      anchor = center,
      width = 4.5cm,
      axis equal,
      axis lines = none,
      axis line style = {thick, gray},
      xmin = -0.2, xmax = 0.2,
      ymin = -1, ymax = 1,
      ticks = none
      ]
      \draw[gray, ->] (axis cs: -0.9,0) -- (axis cs: 1,0) node[below, xshift = -2mm] {\(x\)};
      \draw[gray, ->] (axis cs: 0,-0.9) -- (axis cs: 0,1) node[left, yshift = -2mm] {\(y\)};
      \draw[plotcol1, very thick] (axis cs: 0,0) coordinate (b) -- ++(45:1.2cm) coordinate (a);
      \draw[plotcol1, very thick] (axis cs: 0,0) -- ++(45:-1.2cm);
    \end{axis}
    \node at (0.1, -1.3) {\footnotesize Input polarization};
  \end{scope}
  \draw[dashed] (2.25,0) -- (2.25,0.9);
  \draw[dashed, ->] (2.25,0.9) -- (5.8,0.9);
  \begin{scope}[shift={(6.5,0.5)}, scale=0.75]
    \begin{axis}[
      anchor = center,
      width = 4.5cm,
      axis equal,
      axis lines = none,
      axis line style = {thick, gray},
      xmin = -0.2, xmax = 0.2,
      ymin = -1, ymax = 1,
      ticks = none
      ]
      \draw[gray, ->] (axis cs: -0.9,0) -- (axis cs: 1,0);
      \draw[gray, ->] (axis cs: 0,-0.9) -- (axis cs: 0,1);
      \draw[plotcol1, very thick, opacity=0.5] (axis cs: 0,0) coordinate (b) -- ++(45:1.2cm) coordinate (a);
      \draw[plotcol1, very thick, opacity=0.5] (axis cs: 0,0) -- ++(45:-1.2cm);
      \draw[plotcol2, very thick, rotate around={30:(0,0)}] (0,0) ellipse (0.25cm
      and 1.055cm);
      \draw[<->, dashed] (axis cs: 0,0) -- ++(120:1.055cm) coordinate (c);
      \draw[draw=none] (a) -- (b) -- (c) pic[draw=black, <->, angle
      eccentricity=1.25, angle radius = 0.7cm] {angle=a--b--c};
      \node at (0.2,0.78) {\(\psi\)};
    \end{axis}
    \node at (0.1, -1.3) {\footnotesize Output polarization};    
  \end{scope}

  \draw (9.38, 1.3) -- (9.38, 1.3);

  \node at (-8.5, 1.5) {\textbf{(\thesubfigure)}};  
\end{tikzpicture}} } \hfill
  \subfloat{\label{fig2:trans-spectrum} \parbox{0.42\textwidth}{
      \figswitch{\begin{tikzpicture}
  \begin{axis}[
    width = \linewidth,      
    xlabel= Wavelength, 
    ylabel= Transmittance,
    x unit=\si{\micro\meter}, 
    xmin = 10.75, xmax = 14,
    ymin = 0, ymax = 1.0,
    title= {$d = \SI{1.39}{\micro\meter}$},
    legend entries={\(T_{\parallel}\), \(T_{\perp}\)},
    legend pos = north west
    ]
    \coordinate (inset) at (rel axis cs: 0.95, 0.95);
    \addplot+[very thick] table[col sep=comma,  x=wavelength, y=parallel_T]
    {data/fig2-trans-conversion-spectra-new.csv}; 
    \addplot+[very thick] table[col sep=comma, x=wavelength, y=crossed_T]
    {data/fig2-trans-conversion-spectra-new.csv}; 
    \addplot+[only marks, mark options={draw=black, line width=0.5pt}] table[col
    sep=comma] {data/fig2-trans-conversion-maximum-new.csv};
  \end{axis}
  \node at (-1.1,5.275) {\textbf{(\thesubfigure)}};
  \begin{axis}[
    at={(inset)},
    anchor={north east},
    width=0.4\linewidth,
    height=0.35\linewidth,
    xmin=10, ymax = 1.5,
    xtick = {11, 14},
    xticklabels={\SI{0}{\micro\meter}, \SI{3}{\micro\meter}},
    ytick = {0,0.75,1.5},
    yticklabels={\SI{0}{\micro\meter}, \SI{0.75}{\micro\meter}, \SI{1.5}{\micro\meter}}
    ]
    \addplot+[black, very thick] table[col sep=comma, x=position, y=height] {data/fig2-trans-AFM-profile.csv};
    \draw[<->] (11, 0.1) -- (11, 1.4) node[rotate=90,midway, below, yshift=0.5mm] {\footnotesize \(d\)};
  \end{axis}
\end{tikzpicture}} } } \hfill
  \subfloat{\label{fig2:trans-heatmap} \parbox{0.42\textwidth}{
      \figswitch{\begin{tikzpicture}
  \begin{axis}[       
    width = \linewidth,
    height= 0.565\linewidth,
    enlargelimits=false,
    axis on top,
    ylabel=Thickness \(d\),
    y unit=\si{\micro\meter},
    xmin = 12.45, xmax = 13.15,
    ymin = 0.5, ymax = 2.1,
    xticklabels={},
    title=Transmission,
    legend entries={Linear pol., Experiment},
    legend style = {text = white},
    legend pos=north west,
    name=thickness plot,
    ]
    \addplot+[forget plot] graphics[xmin = 12.35, xmax = 13.15, ymin = 0.19, ymax = 2.1]
    {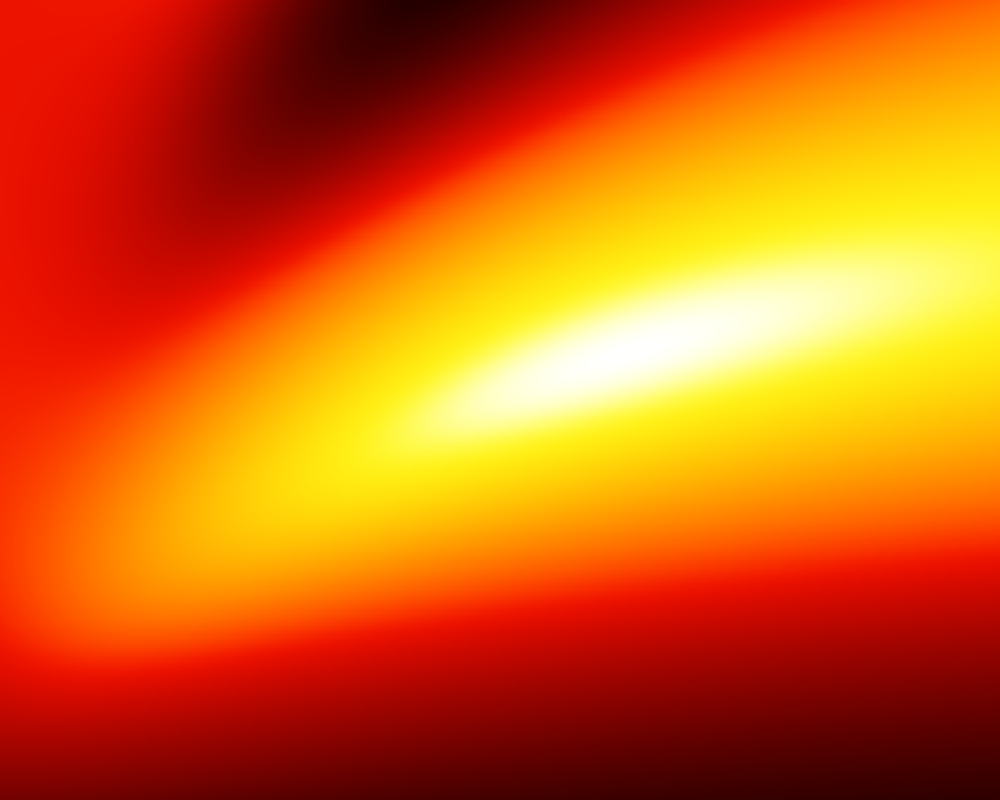};
    \addplot+[forget plot, thin, black, dashed] table[col sep=comma] {data/fig3-trans-psi-contour.csv};
    \addplot+[forget plot, thin, black, dashed] table[col sep=comma] {data/fig3-trans-psi-contour-2.csv};
    \addplot+[plotcol1!50!white, very thick] table[col sep=comma] {data/fig3-trans-max-e.csv};
    \addplot+[plotcol3, only marks, very thick, mark options={draw=black, line width=0.5pt},
    ]
    table[col sep=comma] {data/fig3-trans-conversion-wl-vs-thickness-new.csv};

    \draw[plotcol1, very thick, ->] (12.9526, 1.65) -- (12.9526, 1.45);
  \end{axis}
  \coordinate (plot2) at ($(thickness plot.south) + (0,-0.3)$);
  \begin{axis}[
    width = \linewidth,
    height = 0.46\linewidth,
    ylabel= Conv. ratio,
    xlabel=Wavelength, 
    x unit=\si{\micro\meter}, 
    xmin = 12.45, xmax = 13.15,
    ymin = 0, ymax = 1,
    ytick = {0, 0.5, 1},
    at=(plot2),
    anchor=north,
    ]
    \addplot+[plotcol3, only marks,
    mark options={draw=black, line  width=0.5pt}
    ]
    plot table[col sep=comma, x= wavelength, y = conversion_ratio] {data/fig3-trans-wl-vs-conversion-ratio.csv}; 
  \end{axis}
  \node at (-1.25,3.05) {\textbf{(\thesubfigure)}};
  \node at (-1.25,-0.275) {\textbf{\refstepcounter{subfigure}(\thesubfigure)}};
    

\end{tikzpicture}}
      \refstepcounter{subfigure}\label{fig2:trans-conversion}}}
  \hspace*{-1.250em} \subfloat{ \parbox{0.1\textwidth}{
      \figswitch{\raisebox{2.85cm}{
          \begin{tikzpicture}
            \node[draw,thick,inner sep = 0pt] (bar) at (0,0)
            {\includegraphics[angle=90,
              height=2.5cm,width=0.2cm]{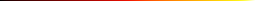}};
            \node[right] at (bar.south east) {0}; \node[right] at (bar.north
            east) {1}; \node[below, rotate=90, yshift=-2mm, align=center] at
            (bar) {Conv. ratio};
          \end{tikzpicture}
        }} } } {} \caption{\label{fig2:full} \textbf{Transmittance via
      \acs{FTIR} micro-spectroscopy.} \textbf{(a)}~Schematic of the beam path of
    the \acs{FTIR} microscope and sample in transmission mode, including
    sketches of the polarization state of the incoming and outgoing
    light. \textbf{(b)}~Transmittance spectra of \acs{MoO} on a Si
    substrate. \(T_{\parallel}\) and \(T_{\perp}\) represent the transmittance
    with parallel and perpendicular output polarizer, respectively, where both
    are normalized to the transmittance of the bare silicon substrate. Inset
    shows the height profile of the flake measured using
    \acs{AFM}. \textbf{(c)}~\acs{MoO} flake thickness \(d\) plotted against
    maximum conversion wavelength. The data point obtained from the spectrum in
    \autoref{fig2:trans-spectrum} is highlighted with the arrow. The background
    color map shows \acs{TMM} calculations of the \acs{ER}, the blue curve
    represents linear-to-linear conversion \(e=1\), the dashed black contours
    mark areas where \(\psi\) is within \(\pm\SI{5}{\degree}\) (inner contour
    line) and \(\pm\SI{10}{\degree}\) (outer contour line) from
    \(\psi = \SI{90}{\degree}\), and the green dots are experimental
    data. \textbf{(d)}~Highest conversion ratio extracted from the same flakes
    as in \textbf{(c)}, with thicknesses shown in the upper panel.}
\end{figure*}
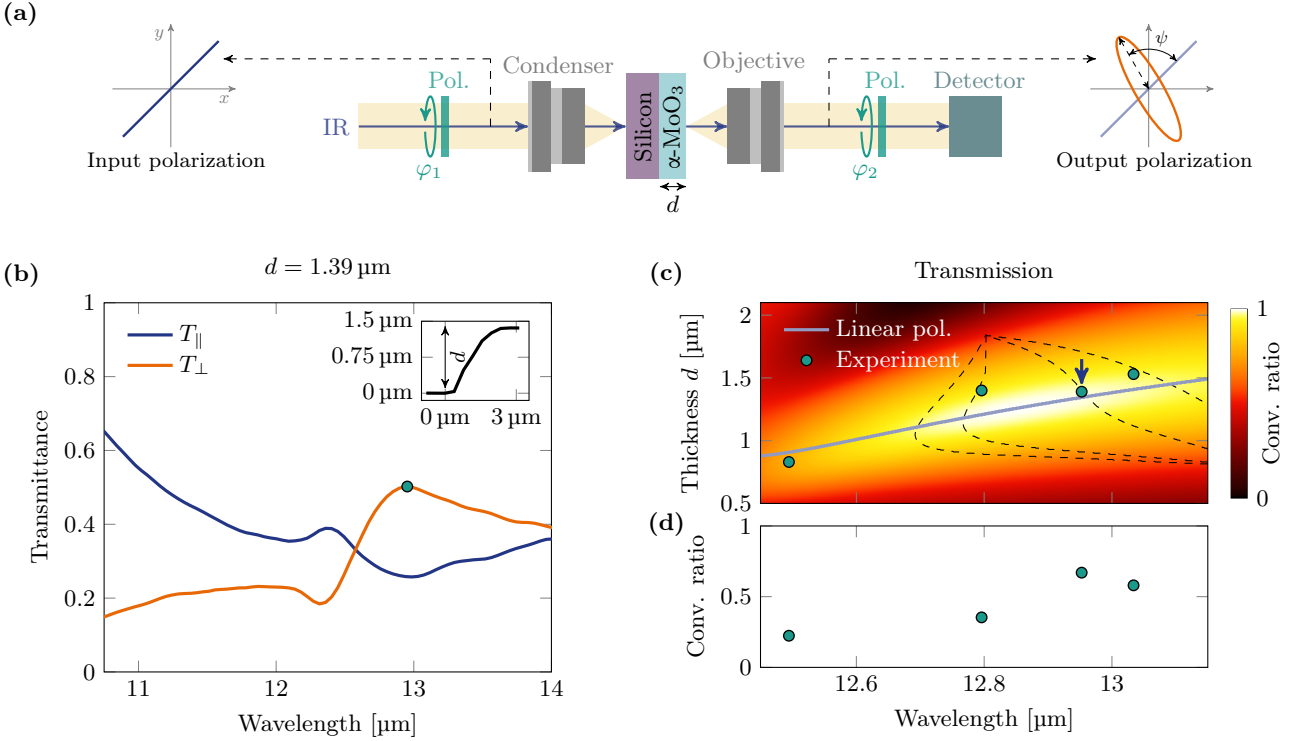

\section{Results}
    
We mechanically exfoliate flakes of \ac{MoO} from bulk crystals and transfer
them to silicon and gold-coated glass substrates in order to realize phase
retarders that operate in transmission and reflection mode,
respectively. Exfoliated flakes of \ac{MoO} typically possess a rectangular
shape, due to the crystal's orthorhombic structure~\cite{alvarez-perez2020am}
(\autoref{fig1:permittivity}a). We refer to the crystal direction \([100]\) as
the \(x\)-axis, to the \([001]\) as the \(y\)-axis, and to the \([010]\)
direction as the \(z\)-axis. We measure the dielectric permittivity of \ac{MoO}
using \ac{FTIR} spectroscopy, implementing the method described
in~\cite{sarkar2023}. The results we obtain, shown in
\autoref{fig1:permittivity}b and \autoref{fig1:permittivity}c, are in agreement
with previous reports~\cite{ma2018n, zheng2018am, alvarez-perez2020am}. As shown
in panel b, near \SI{12}{\micro\meter}, the dielectric function along the
\(x\)-axis ($\epsilon_\mathrm{x}$) resonates due to a \ac{TO} phonon occurring
at the corresponding photon energy, while $\epsilon_\mathrm{y}$ remains very
small. Thus, the birefringence of \ac{MoO} in this spectral region reaches
values near 15.
 
Aiming to achieve $90$ degrees polarization rotation, we note that, while
\autoref{eq:half-wave-plate} appropriately describes the operation of optically
thick and lossless dielectric wave plates, the class of \ac{vdW} materials
considered here requires taking into account their strong frequency dispersion
due to phonon resonances as well as the non-negligible dichroism
near-resonance. The latter makes orthogonal components of the electromagnetic
fields experience different level of losses, whereas the former yields strong
sensitivity to the optimal operational wavelength with respect to small changes
in a flake's thickness. Another important factor is the extremely small
thickness of the phase retarders as compared to the wavelength. In this deeply
subwavelength region, interference effects take precedence in describing the
conversion process. It is thus essential to consider not only the phase acquired
upon propagation within the slab, but also the reflection and transmission at
its boundaries~\cite{holmes1964josaj}. In the following sections, we utilize
dichroism to effectively leverage constructive interference so that light
maximally experiences the anisotropy of the crystal. Through a combination of
simulations and experiments, we illustrate our approach to achieve significant
polarization conversion while maintaining low insertion losses.
 
\begin{figure*}[t]
  \flushleft \subfloat{\label{fig3:setup}
    \figswitch{\begin{tikzpicture}
  \begin{scope}[xshift=2.5cm, yshift=-2mm]
    \fill[plotcol4!85!plotcol2, opacity=0.7] (-5, -0.3) rectangle (-1.7, 0.3);
    \fill[plotcol4!85!plotcol2, opacity=0.7] (-1, -0.3) -- (-0.45, 0) -- (-1, 0.3) -- cycle;
    \fill[plotcol4!85!plotcol2, opacity=0.7] (-3.05, 0.3) rectangle (-2.45, 1.5);    
    
    \draw[<->, thick, plotcol1, opacity=0.8] (-5,0) -- (-1.75,0);
    \draw[<->, thick, plotcol1, opacity=0.8] (-1,0) -- (-0.45,0);
    \draw[<-, thick, plotcol1, opacity=0.8] (-2.75, 0) -- (-2.75, 1.5) node[above] {IR};

    \draw[very thick, dashed, opacity=0.8] ($(-2.75, 0) + (135:0.05)$) -- ++(135:0.5);
    \draw[very thick, dashed, opacity=0.8] (-2.75, 0) -- ++(135:-0.5);

    \fill[shadecol2] (-0.1,-0.7) rectangle (-0.45,0.7) node[midway, black, rotate=90] {\acs{MoO}};
    \fill[gold] (-0.1,-0.7) rectangle (0.2,0.7) node[midway, black, rotate=90] {Gold};
    \fill[shadecol1] (0.2,-0.7) rectangle (0.6,0.7) node[midway, black, rotate=90] {Glass};  
    \draw[<->] (-0.1, 0.8) -- (-0.45, 0.8) node[midway, above] {\(d\)};

    \fill[gray] (-1.5, 0.5) rectangle (-1, -0.5);
    \fill[gray!50!white] (-1.45, 0.5) rectangle (-1.3, -0.5);
    \fill[gray] (-1.45, 0.6) node[above, xshift=1mm] {Objective} rectangle (-1.75, -0.6);
    \fill[gray!50!white] (-1.75, -0.6) rectangle (-1.7, 0.6);

    \fill[plotcol5] (-5.7, 0.4) rectangle (-5, -0.4) node[below, xshift=-3mm] {Detector};

    \begin{scope}[yshift=4.5mm]
    \fill[plotcol3, opacity=0.8] (-2.35, 0.6) rectangle (-3.15, 0.7) node[left] {Pol.};
    \draw[thick, plotcol3, ->] (-2.75,0.85) [partial ellipse=-260:80:0.4cm and
    0.07cm] node[pos=0.8, right] {\(\varphi_{1}\)};
    \end{scope}
  
    \fill[plotcol3, opacity=0.8] (-3.8, 0.4) node[above] {Pol.} rectangle (-3.9, -0.4);
    \draw[thick, plotcol3, ->] (-4.05,0) [partial ellipse=-170:170:0.07cm and
    0.4cm] node[pos=0.2, below] {\(\varphi_{2}\)}; 
  \end{scope}

  \draw[dashed, ->] (-0.25,0.65) -- (-5.5,0.65);
  \begin{scope}[shift={(-6.25,0.25)}, scale=0.75]
    \begin{axis}[
      anchor = center,
      width = 4.5cm,
      axis equal,
      axis lines = none,
      axis line style = {thick, gray},
      xmin = -0.2, xmax = 0.2,
      ymin = -1, ymax = 1,
      ticks = none
      ]
      \draw[gray, ->] (axis cs: -0.9,0) -- (axis cs: 1,0) node[below, xshift = -2mm] {\(x\)};
      \draw[gray, ->] (axis cs: 0,-0.9) -- (axis cs: 0,1) node[left, yshift = -2mm] {\(y\)};
      \draw[plotcol1, very thick] (axis cs: 0,0) coordinate (b) -- ++(45:1.2cm) coordinate (a);
      \draw[plotcol1, very thick] (axis cs: 0,0) -- ++(45:-1.2cm);
    \end{axis}
    \node at (0.1, -1.3) {\footnotesize Input polarization};
  \end{scope}
  \draw[dashed] (-1,-0.2) -- (-1,-1.1) -- (3.85,-1.1) -- (3.85,0.65);
  \draw[dashed, ->] (3.85,0.65) -- (5.2,0.65);
  \begin{scope}[shift={(5.75,0.25)}, scale=0.75]
    \begin{axis}[
      anchor = center,
      width = 4.5cm,
      axis equal,
      axis lines = none,
      axis line style = {thick, gray},
      xmin = -0.2, xmax = 0.2,
      ymin = -1, ymax = 1,
      ticks = none
      ]
      \draw[gray, ->] (axis cs: -0.9,0) -- (axis cs: 1,0);
      \draw[gray, ->] (axis cs: 0,-0.9) -- (axis cs: 0,1);
      \draw[plotcol1, very thick, opacity=0.5] (axis cs: 0,0) coordinate (b) -- ++(45:1.2cm) coordinate (a);
      \draw[plotcol1, very thick, opacity=0.5] (axis cs: 0,0) -- ++(45:-1.2cm);
      \draw[plotcol2, very thick, rotate around={30:(0,0)}] (0,0) ellipse (0.25cm
      and 1.055cm);
      \draw[<->, dashed] (axis cs: 0,0) -- ++(120:1.055cm) coordinate (c);
      \draw[draw=none] (a) -- (b) -- (c) pic[draw=black, <->, angle
      eccentricity=1.25, angle radius = 0.7cm] {angle=a--b--c};
      \node at (0.2,0.78) {\(\psi\)};
    \end{axis}
    \node at (0.1, -1.3) {\footnotesize Output polarization};
  \end{scope}
  
  \draw (9.38, 1.3) -- (9.38, 1.3);

  \node at (-8.5, 1.5) {\textbf{(\thesubfigure)}};  
\end{tikzpicture}} } \hfill
  \subfloat{\label{fig3:ref-spectrum} \parbox{0.42\textwidth}{
      \figswitch{\begin{tikzpicture}
  \begin{axis}[
    width = \linewidth,      
    xlabel= Wavelength, 
    ylabel= Reflectance,
    x unit=\si{\micro\meter}, 
    xmin = 9.75, xmax = 16,
    ymin = 0, ymax = 1,
    title={\(d = \SI{920}{\nano\meter}\)},
    legend entries={\(R_{\parallel}\), \(R_{\perp}\)},
    legend style = {at = {(0.05,0.5)}, anchor=west},
    ]
    \coordinate (inset) at (rel axis cs: 0.93, 0.95);

    \addplot+[very thick] table[col sep=comma,  x expr=\thisrow{wavelength}
    , y expr=\thisrow{parallel_T}] {data/fig2-ref-conversion-spectra.csv};
    \addplot+[very thick] table[col sep=comma, x expr=\thisrow{wavelength},
    y expr=\thisrow{crossed_T}] {data/fig2-ref-conversion-spectra.csv};
    \addplot+[only marks, mark options={draw=black, line width=0.5pt}] table[col sep=comma]
    {data/fig2-ref-conversion-maximum.csv};        
  \end{axis}
  \node at (-1.1,5.275) {\textbf{(\thesubfigure)}}; 
  \begin{axis}[
    at={(inset)},
    anchor={north east},
    width=0.4\linewidth,
    height=0.35\linewidth,
    ymax = 1,
    xtick = {4,8},
    xticklabels={\SI{0}{\micro\meter}, \SI{4}{\micro\meter} },
    yticklabels={ ,\SI{0}{\micro\meter}, \SI{0.5}{\micro\meter}, \SI{1}{\micro\meter}}
    ]
    \addplot+[black, very thick] table[col sep=comma, x=position, y=height] {data/fig2-ref-AFM-profile.csv};
    \draw[<->] (5, 0.1) -- (5, 0.85) node[rotate=90,midway, above, yshift=-0.5mm] {\footnotesize \(d\)};
  \end{axis}
\end{tikzpicture}} } } \hfill
  \subfloat{\label{fig3:ref-heatmap} \parbox{0.42\textwidth}{
      \figswitch{\begin{tikzpicture}
  \begin{axis}[       
    width = \linewidth,
    height = 0.565\linewidth,
    enlargelimits=false,
    axis on top,        
    ylabel=Thickness \(d\),
    y unit=\si{\micro\meter},
    xmin = 12.4, xmax = 15.1,
    ymin = 0.19, ymax = 1.9,
    xticklabels={},
    title=Reflection,
    legend entries={Linear pol., Experiment},
    legend style = {text = white},        
    legend pos=north west,
    name=thickness plot
    ]
    \addplot+[forget plot] graphics[xmin = 12.4, xmax = 15.1, ymin = 0.19, ymax = 2.1] {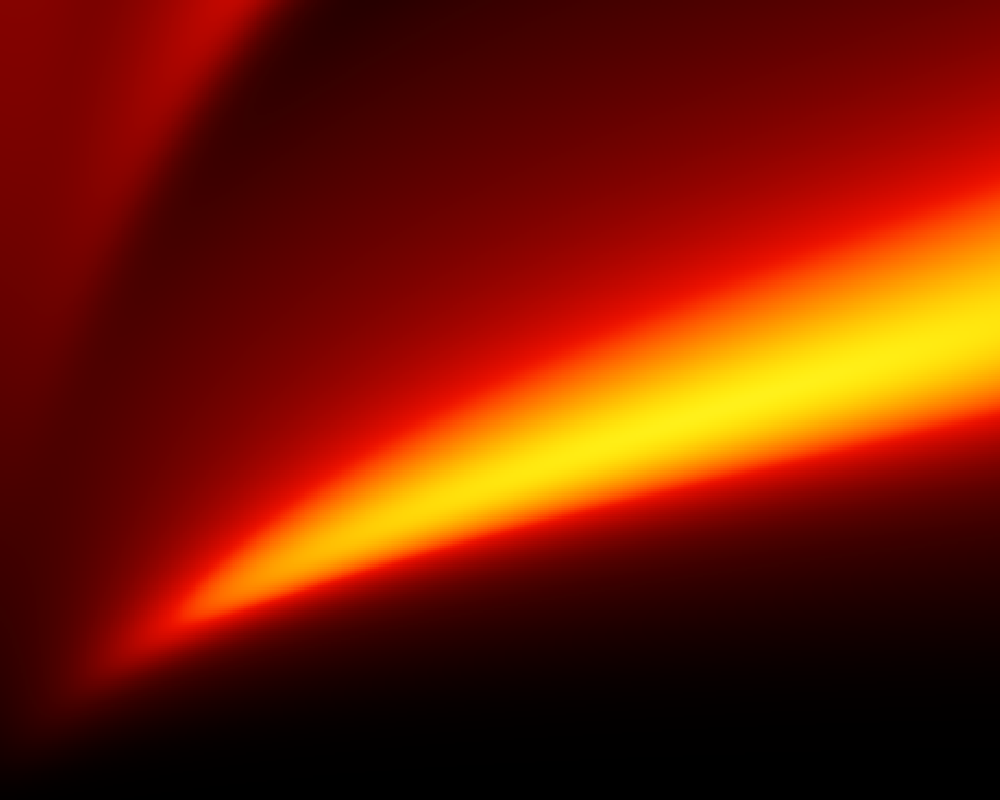};
    \addplot+[forget plot, thin, white, dashed] table[col sep=comma] {data/fig3-ref-psi-contour.csv};
    \addplot+[forget plot, thin, white, dashed] table[col sep=comma] {data/fig3-ref-psi-contour-2.csv};
    \addplot+[plotcol1!50!white, very thick] table[col sep=comma] {data/fig3-ref-max-e.csv};
    \addplot+[plotcol3, only marks, very thick, mark options={draw=black, line width=0.5pt},
    ]
    table[col sep=comma] {data/fig3-ref-conversion-wl-vs-thickness.csv};

    \draw[very thick, ->, plotcol1] (13.9184, 1.2) -- (13.9184, 1);
  \end{axis}
  \coordinate (plot2) at ($(thickness plot.south) + (0,-0.3)$);
  \begin{axis}[
    width = \linewidth,
    height = 0.46\linewidth,
    ylabel= Conv. ratio,
    xlabel=Wavelength, 
    x unit=\si{\micro\meter}, 
    xmin = 12.4, xmax = 15.1,
    at=(plot2),
    anchor=north,
    ytick = {0, 0.5, 1},
    ymin = 0, ymax = 1,
    ]
    \addplot+[plotcol3, only marks, very thick, mark options={draw=black, line
      width=0.5pt}
    ]
    plot table[col sep=comma, x=wavelength, y = conversion_ratio] {data/fig3-ref-wl-vs-conversion-ratio.csv};
  \end{axis}
  \node at (-1.25,3.05) {\textbf{(\thesubfigure)}};
  \node at (-1.25,-0.275) {\textbf{\refstepcounter{subfigure}(\thesubfigure)}};
  

\end{tikzpicture}}
      \refstepcounter{subfigure}\label{fig3:ref-conversion}}}
  \hspace*{-1.250em} \subfloat{ \parbox{0.1\textwidth}{ \figswitch{
        \raisebox{2.85cm}{
          \begin{tikzpicture}
            \node[draw,thick,inner sep = 0pt] (bar) at (0,0)
            {\includegraphics[angle=90,
              height=2.5cm,width=0.2cm]{data/fig3-trans-simulation-colorbar.png}};
            \node[right] at (bar.south east) {0}; \node[right] at (bar.north
            east) {1}; \node[below, rotate=90, yshift=-2mm, align=center] at
            (bar) {Conv. ratio};
          \end{tikzpicture}
        }} } } \hfill{}
  \caption{\label{fig3:full}\textbf{Reflectance via \acs{FTIR}
      micro-spectroscopy.} \textbf{(a)}~Schematic showing the beam path of the
    \acs{FTIR} microscope and the sample in reflection mode and sketches of the
    polarization state of the incoming and outgoing light. \textbf{(b)},
    \textbf{(c)}, and \textbf{(d)} show the reflectance analogues of the
    corresponding panels in \autoref{fig2:full}, except that the dashed black
    contours in \textbf{c} mark areas where \(\psi\) is within
    \(\pm\SI{5}{\degree}\) (inner contour line) and \(\pm\SI{10}{\degree}\)
    (outer contour line) from \(\psi = \SI{90}{\degree}\).}
\end{figure*}

The topography of exfoliated flakes is characterized with \ac{AFM}, while their
spectral transmittance and reflectance is measured with \ac{FTIR}
micro-spectroscopy. To characterize the flakes, we control the input beam's
polarization and read the output beam's polarization using two wire-grid
polarizers, in both transmission and reflection modes, as shown in
\autoref{fig2:setup} and \autoref{fig3:setup}, respectively. The input polarizer
is oriented at \SI{45}{\degree} with respect to the crystal's axes, as shown
with the blue straight line in the sketches next to the setup in
\autoref{fig2:setup} and \autoref{fig3:setup}. This is a sketch of the
polarization ellipse of the input beam with respect to the two crystallographic
axes. As a result of the input polarizer, the input polarization ellipse reduces
to a straight line, oriented at an angle of \SI{45}{\degree} with respect to the
\(x\)-axis. Correspondingly, with the orange shape in the sketch next to the
output polarizer in \autoref{fig2:setup} and \autoref{fig3:setup}, we represent
a general elliptical polarization state with arbitrary ellipticity, defined by
the quantity \(e= \sqrt{1 - \nfrac{b^{2}}{a^{2}}}\), with \(a\) and \(b\) being
the half-lengths of the major and minor axes of the ellipse, respectively. For
linearly polarized output light, the ellipse collapses into a line \(e = 1\),
whereas circularly polarized light traces a circle \(e = 0\), corresponding to
circular polarization. In the same schematics of \autoref{fig2:setup} and
\autoref{fig3:setup}, we denote \(\psi\), which is the rotation angle of the
principal axis of the output beam's polarization ellipse, with respect to the
input polarization, termed henceforth polarization rotation. In the case of a
lossless dielectric crystal, the representation of a phase retarder via the pair
(\(e\), \(\psi\)) reduces to the standard description via the retardance
\(\delta\) (\(\delta=\pi\) and \(\delta=\pi/2\) correspond to a half-wave plate
and quarter-wave plate operation, respectively), quoted in commercial wave
plates.

We use simulations of reflectance and transmittance, performed using an in-house
implementation of the \ac{TMM}~\cite{passler2017josab,enders2023}, to estimate
the optimal flake thickness for which the polarization rotation approaches $90$
degrees in the vicinity of \SI{12}{\micro\meter}, where birefringence is
maximal. As shown in \autoref{fig2:trans-heatmap} and
\autoref{fig3:ref-heatmap}, for both transmission and reflection, this occurs
for thicknesses near \SI{1}{\micro\meter}. This thickness yields constructive
interference within the flake, thus light maximally experiences the anisotropy
of the crystal resulting in large linear polarization rotation while suppressing
multiple reflections~\cite{Pietraszkiewicz1995}. In transmission mode
(\autoref{fig2:full}), we define the polarization conversion ratio as
\(\nfrac{T_{\perp}}{\left(T_{\parallel} + T_{\perp}\right)}\) and the phase
retarder's \ac{IL} as \(-10\log(T_{\perp}+T_{\parallel})\), where \(T_{\perp}\)
and \(T_{\parallel}\) are the measured transmittance with the output polarizer
perpendicular and parallel to the input one, respectively. Both \(T_{\perp}\)
and \(T_{\parallel}\) are normalized to \(T_{\parallel,sub}\), taken on the bare
silicon substrate. In this notation, therefore, \(T_{\perp}\) denotes the
absolute conversion. \autoref{fig2:trans-spectrum} shows the transmittance
spectra measured for a flake of thickness \SI{1.39}{\micro\meter} (inset shows
the \ac{AFM} data). As can be seen from \(T_{\perp}\), maximum polarization rotation
occurs at \SI{12.95}{\micro\meter}, and the absolute conversion is approximately
\SI{50}{\percent}.

Using the \ac{TMM}, we show in \autoref{fig2:trans-heatmap} a color map
representing the calculated polarization conversion ratio as a function of
flake's thickness and wavelength. To gain a better understanding of the position
of the maximum conversion ratio, we extract the ellipticity \(e\) and the
polarization rotation angle \(\psi\) for each wavelength and thickness. The solid blue curve in
\autoref{fig2:trans-heatmap} shows the points where the output polarization is
purely linear, meaning \(e=1\). Evidently, the maximum conversion ratio is
achieved when \(\psi = \SI{90}{\degree}\) along this line. The dashed black
contours mark the areas where \(\psi\) is within \(\pm\SI{5}{\degree}\) (inner
contour line) and \(\pm\SI{10}{\degree}\) (outer contour line) from
\(\psi = \SI{90}{\degree}\), hence the intersection between the blue curve and
the dashed contours represents near-ideal quarter-wave plate
operation. Experimental results are superimposed and shown as green dots, each
corresponding to a particular flake with given thickness (\(x\)-axis). The
wavelength at which each measured data point is placed corresponds to the
wavelength at which optimal polarization conversion (\(y\)-axis) was
measured. We observe good agreement between the experimental results and the
solid line representing purely linear polarization (half-wave plate
operation). Notably, three of our measurements, corresponding to three different
flakes with different thicknesses, fall within a polarization rotation angle
range between \SIrange{80}{100}{\degree} with high conversion ratios. From
\autoref{fig2:trans-conversion}, the maximum experimentally measured
polarization conversion ratio is \num{0.67}, and is achieved for
\(d\approx\SI{1.4}{\micro\meter}\) for the wavelength of \SI{12.95}{\micro\meter} as
shown by the vertical arrow. For all the experimental data points, the \ac{IL}
remains below \SI{1.3}{\decibel}. This was expected from the discussion above,
owning to the deeply subwavelength thickness of the flakes, compensating for the
large $\mathrm{Im}(\epsilon_\mathrm{x})$ (\autoref{fig1:permittivity}c).

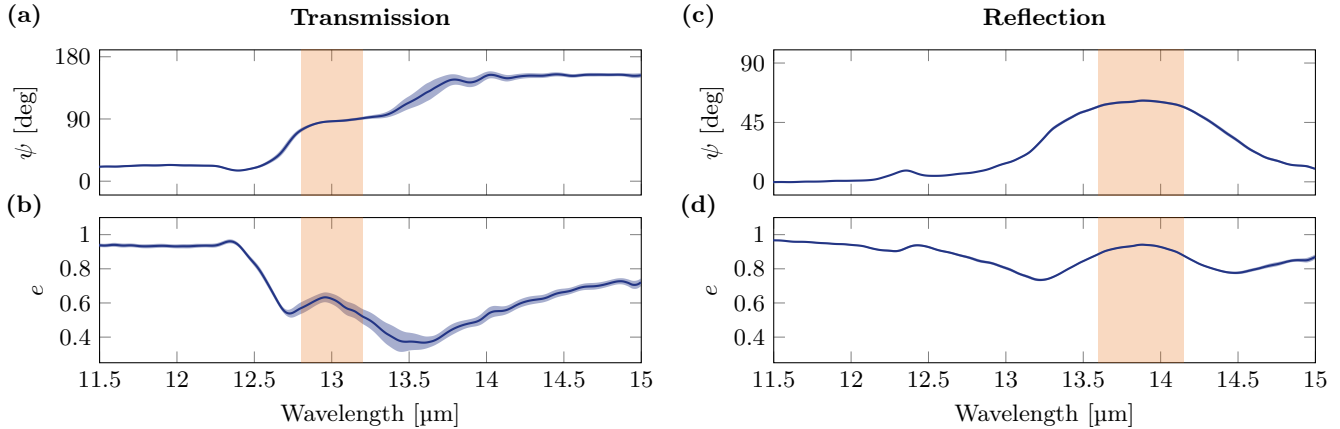
\begin{figure*}[t]
  \subfloat{\label{fig4:angle-trans} \parbox{0.49\textwidth}{
      \figswitch{\begin{tikzpicture}
  \begin{axis}[       
    width = 1\linewidth,
    height= 0.4\linewidth,
    enlargelimits=false,
    axis on top,
    ylabel= {\(\psi\)},
    y unit=\si{deg}, 
    title = \textbf{Transmission},
    xmin = 11.5, xmax = 15,
    ymin = -20, ymax = 190,
    xticklabels={},
    ytick={0, 90, 180},
    name=plot,
    ]
    \fill[opacity=0.25, plotcol2] (12.8,-30) rectangle (13.2,195);
    \addplot [name path=psi trans upper, draw=none, forget plot]
             table[col sep = comma, x=wavelength, y expr=\thisrow{psi}+\thisrow{psi_err}]
             {data/fig5-trans-spec.csv};
    \addplot [name path=psi trans lower, draw=none, forget plot]
             table[col sep = comma, x=wavelength, y expr=\thisrow{psi}-\thisrow{psi_err}]
             {data/fig5-trans-spec.csv};
    \addplot [fill=plotcol1, opacity=0.4, forget plot] fill between[of=psi trans upper and psi trans lower];
    \addplot+[plotcol1, thick] table[col sep=comma, x=wavelength, y=psi] {data/fig5-trans-spec.csv};
  \end{axis}
  \coordinate (plot2) at ($(plot.south) + (0,-0.3)$);
  \begin{axis}[
    width = 1\linewidth,
    height = 0.4\linewidth,
    ylabel= {\(e\)},
    xlabel= Wavelength, 
    x unit= \si{\micro\meter}, 
    xmin = 11.5, xmax = 15,
    ymax = 1.1, ymin = 0.25,
    at=(plot2),
    anchor=north,
    ]
    \fill[opacity=0.25, plotcol2] (12.8,0.2) rectangle (13.2,1.2);
    \addplot [name path=e trans upper, draw=none, forget plot]
             table[col sep = comma, x=wavelength, y expr=\thisrow{e}+\thisrow{e_err}]
             {data/fig5-trans-spec.csv};
    \addplot [name path=e trans lower, draw=none, forget plot]
             table[col sep = comma, x=wavelength, y expr=\thisrow{e}-\thisrow{e_err}]
             {data/fig5-trans-spec.csv};
    \addplot [fill=plotcol1, opacity=0.4, forget plot] fill between[of=e trans upper and e trans lower];
    \addplot+[plotcol1, thick] table[col sep=comma, x=wavelength, y=e] {data/fig5-trans-spec.csv};
  \end{axis}
  \node at (-1,2.375) {\textbf{(\thesubfigure)}};
  \node at (-1,-0.15) {\refstepcounter{subfigure}\textbf{(\thesubfigure)}};
\end{tikzpicture}}
      \refstepcounter{subfigure}\label{fig4:ellipticity-trans}}}\hfill
  \subfloat{\label{fig4:angle-ref} \parbox{0.49\textwidth}{
      \figswitch{\begin{tikzpicture}
  \begin{axis}[       
    width = 1\linewidth,
    height= 0.4\linewidth,
    enlargelimits=false,
    axis on top,
    ylabel= {\(\psi\)},
    y unit=\si{deg}, 
    title = \textbf{Reflection},
    xmin = 11.5, xmax = 15,
    ymin = -10, ymax = 100,
    xticklabels={},
    ytick={0, 45, 90},
    name=plot,
    ]
    \fill[opacity=0.25, plotcol2] (13.6,-30) rectangle (14.15,195);
    \addplot [name path=psi ref upper, draw=none, forget plot]
             table[col sep = comma, x=wavelength, y expr=\thisrow{psi}+\thisrow{psi_err}]
             {data/fig5-ref-spec.csv};
    \addplot [name path=psi ref lower, draw=none, forget plot]
             table[col sep = comma, x=wavelength, y expr=\thisrow{psi}-\thisrow{psi_err}]
             {data/fig5-ref-spec.csv};
    \addplot [fill=plotcol1, opacity=0.4, forget plot] fill between[of=psi ref upper and psi ref lower];
    \addplot+[plotcol1, thick] table[col sep=comma, x=wavelength, y=psi] {data/fig5-ref-spec.csv};
  \end{axis}
  \coordinate (plot2) at ($(plot.south) + (0,-0.3)$);
  \begin{axis}[
    width = \linewidth,
    height = 0.4\linewidth,
    ylabel= {\(e\)},
    xlabel= Wavelength, 
    x unit= \si{\micro\meter}, 
    xmin = 11.5, xmax = 15,
    ymax = 1.1, ymin = 0.25,
    at=(plot2),
    anchor=north,
    ]
    \fill[opacity=0.25, plotcol2] (13.6,0.2) rectangle (14.15,1.2);
    \addplot [name path=e ref upper, draw=none, forget plot]
             table[col sep = comma, x=wavelength, y expr=\thisrow{e}+\thisrow{e_err}]
             {data/fig5-ref-spec.csv};
    \addplot [name path=e ref lower, draw=none, forget plot]
             table[col sep = comma, x=wavelength, y expr=\thisrow{e}-\thisrow{e_err}]
             {data/fig5-ref-spec.csv};
    \addplot [fill=plotcol1, opacity=0.4, forget plot] fill between[of=e ref upper and e ref lower];
    \addplot+[plotcol1, thick] table[col sep=comma, x=wavelength, y=e] {data/fig5-ref-spec.csv};
  \end{axis}
  \node at (-1,2.375) {\textbf{(\thesubfigure)}};
  \node at (-1,-0.15) {\refstepcounter{subfigure}\textbf{(\thesubfigure)}};
\end{tikzpicture}}
      \refstepcounter{subfigure}\label{fig4:ellipticity-ref}}}\hfill{}
  \caption{\label{fig4:full}\textbf{Polarization state characterization.}  The
    polarization rotation angle \(\psi\) of the transmitted and reflected beam in
    each mode of operation are reconstructed in \textbf{(a)}~transmission (flake
    thickness: \SI{1.39}{\micro\meter}, silicon substrate) and
    \textbf{(c)}~reflection (flake thickness: \SI{0.92}{\micro\meter}, on
    Au). The ellipticity \(e\) for transmission and reflection is shown in
    \textbf{(b)} and \textbf{(d)}, respectively. The ribbon around the lines
    shows the standard deviation of the numerically fitted parameters. The
    orange shaded region shows the spectral range where the rotation \(\psi\) is close to
    \SI{90}{\degree}.}
\end{figure*}

We repeat the phase retardation characterization in reflection mode
(\autoref{fig3:full}), where \(R_{\parallel}\) and \(R_{\perp}\) correspond to
the output polarizer parallel and perpendicular to the input polarizer,
respectively, as shown in \autoref{fig3:setup}, normalized to the gold-coated
substrate. We present in \autoref{fig3:ref-spectrum} the measured reflectance
spectra for a flake of thickness \SI{920}{\nano\meter} (\ac{AFM} measurement in
inset). For this flake, the maximum absolute conversion (\(R_{\perp}\)) is
\SI{64}{\percent} near \SI{13.9}{\micro\meter}. The maximum conversion is
determined for each flake in the same manner as in transmission mode
(\autoref{fig2:full}), and varies for different flakes and different wavelengths
depending on thickness. The green data points in \autoref{fig3:ref-heatmap}
represent these experimental results, and are superimposed with the
corresponding \ac{TMM} numerical calculations. The solid blue curve corresponds
to pure linear polarization with \(e=1\). Unlike the transmission mode, in
reflection mode, due to additional losses in the Au, the maximum predicted
polarization rotation \(\psi\) deviates from \SI{90}{\degree}, as also verified
experimentally (\autoref{fig4:full}). The maximum polarization rotation angle
predicted is \SI{75}{\degree}. The regions where \(\psi\) exceeds
\SI{70}{\degree} and \SI{60}{\degree} are depicted by the inner and outer black
dashed contour lines in white, respectively. The intersection between the blue
curve and the dashed contours represents operation as close as possible to
quarter-wave plate in reflection mode. Experimentally, although we do not expect
maximal polarization rotation of \SI{90}{\degree}, we still measure considerably
high conversion ratio from one polarization state to its orthogonal
counterpart. In reflection mode, light traverses the flake twice, thus
experiencing the strong birefringence twice. This yields high conversion for a
wider range of wavelengths than in transmission mode, as expected. This is shown
in \autoref{fig3:ref-heatmap}, where one can observe the potential to passively
tune the operational wavelength and achieve linear polarization conversion by
controlling the thickness of the flake within the range of
\SIrange{0.45}{1.5}{\micro\meter}. This enables operation within the wavelength
range of \SIrange{12.6}{15}{\micro\meter}, covering the entire Reststrahlen band
of the \ac{MoO} (\autoref{fig1:permittivity}b), demonstrating that this effect
does not stem from the hyperbolic behavior of \ac{MoO} but rather from the
resonant nature of its birefringence. Finally, as shown in
\autoref{fig3:ref-conversion}, we obtain conversion ratios up to \num{0.82},
with \ac{IL} remaining below \SI{1.7}{\decibel} for all measured flakes.

The results in \autoref{fig2:full} and \autoref{fig3:full} describe the
intensity of the output beam's component that is \num{90} degrees rotated with
respect to the input beam's polarization. Nonetheless, these measurements do not
suffice in characterizing a phase retarder. In particular in the case of
considerable frequency dispersion and absorption (\autoref{fig1:permittivity}b
and \autoref{fig1:permittivity}c), one ought to measure the ellipticity of the
output beam, as linear polarization is not necessarily preserved. To do this
explicitly (while probing depolarization and handedness), the mere use of two
polarizers is insufficient, as one requires an additional phase
retarder. Nonetheless, as mentioned in the introduction, commercial phase
retarders are not available in the \ac{mir} range. Hence, instead, following
basic principles of ellipsometry~\cite{tompkins2005handbook}, we illuminate the
samples with linearly polarized light and measure the outgoing light in two
configurations while rotating the sample: with the two linear polarizers
parallel and orthogonal to each other. Assuming absence of depolarization, as expected in the
deep-subwavelength range that these \ac{MoO} flakes operate, via a numerical fit
to the experimental data, we determine \(\psi\) and \(e\) as a function of the
wavelength~\cite{kilchoer2019ap}. These are shown in \autoref{fig4:angle-trans}
and \autoref{fig4:angle-ref} for transmission and reflection mode, respectively,
whereas the corresponding ellipticity is shown in
\autoref{fig4:ellipticity-trans} and \autoref{fig4:ellipticity-ref}.

As shown in \autoref{fig4:angle-trans}, in transmission mode, the polarization
rotation angle reaches \num{90} degrees for the spectral range from
\SIrange{12.8}{13.2}{\micro\meter}, as was initially designed. The ellipticity
of the transmitted beam (\autoref{fig4:ellipticity-trans}) deviates from the
ideal value of unity for linear polarization, as expected due to
losses. Nonetheless, this can be easily corrected with another polarization
element on top of the flake or by directly etching a wire grid structure into
the sample. In reflection mode, in \autoref{fig4:angle-ref}, we show that the
rotation angle \(\psi\) is approximately \SI{50}{\degree}, as expected from our
simulations shown in \autoref{fig3:ref-heatmap}. In
\autoref{fig4:ellipticity-ref} the ellipticity \(e\) of the reflected beam,
however, is close to unity over the entire wavelength range for which it was
designed, as highlighted in the orange shaded region from
\SIrange{13.6}{14.15}{\micro\meter}. This shows that linear polarization is
preserved while being rotated. The deviations from the ideal behavior of a
half-wave plate operation both both flakes operating in transmission and
reflection mode are easilly corrected with polarizers that do exist in the
mid-IR range.

\section{Discussion}

We design and characterize a wide range of phase retarders composed solely by
exfoliated flakes of \ac{MoO}. By controlling the flakes' thicknesses, we
demonstrate polarization rotation in both transmission and reflection mode, with
conversion ratios as high as \SI{82}{\percent}. Phase retardation is
demonstrated over the spectral range from \SIrange{12}{15}{\micro\meter}, where
bulk crystals perform poorly and commercial phase retarders are scarce. Due to
the deeply subwavelength thickness of the flakes, the absorption of \ac{MoO}
does not considerably affect the performance of the phase retarders, and
insertion losses remain extremely low. Nonetheless, the losses introduce a
degree of circular polarization upon transmission, which, however, can be
corrected by further adjusting the flakes' thickness and using commercial
polarizers.

Our experimental results demonstrate the potential of naturally in-plane
anisotropic \ac{vdW} materials as highly efficient phase retarders. This phase
retardation is not associated with the hyperbolic dispersion of \ac{MoO}, but
rather the result of the resonant features in its dielectric permittivity due to
crystal vibrations that occur at different frequencies along different lattice
directions. Our results open new possibilities for practical applications in the
\ac{mir} spectral range, particularly beyond wavelengths of
\SI{10}{\micro\meter} where conventional, where conventional bulk materials
perform poorly. The principle of operation of the phase retardation scheme
discussed in this work is general and applies to any polar dielectric
low-dimensional anisotropic medium. Thus, these results can be generalized to
other spectral ranges, where other materials exhibit Reststrahlen bands, for
example near \SI{18}{\micro\meter} using \ac{VO}~\cite{taboada-gutierrez2020nm}
or even to \ac{MoO} itself for the resonance along the \(y\)-direction (see
\autoref{fig1:material-comparison}). As crystal growth techniques continue to
improve, emerging \ac{vdW} materials like \ac{MoO} offer a new paradigm as
polarization control elements, beyond the reach of conventional materials for
integrated on-chip photonic devices.

\section{Acknowledgments}
This work is dedicated to the memory of John S. Papadakis. The authors declare
no competing financial interest. The authors would like to thank V. Pruneri for
generously granting access to his laboratory and equipment, as well as engaging
in fruitful discussions. G.T.P. acknowledges financial support from the la Caixa
Foundation (ID 100010434).  M.E. acknowledges financial support from MCIN/AEI/
10.13039/501100011033 (PRE2020-094401) and FSE ``El FSE invierte en tu
futuro''. M. G. acknowledges financial support from the Severo Ochoa Excellence
Fellowship. This work was supported by the Spanish MICINN (PID2021-125441OA-I00,
PID2020–112625GB-I00, and CEX2019-000910- S), the European Union (fellowship
LCF/BQ/PI21/11830019 under the Marie Skłodowska-Curie Grant Agreement
No. 847648), Generalitat de Catalunya (2021 SGR 01443) through the CERCA
program, Fundació Cellex, and Fundació Mir-Puig.  F.H.L.K. acknowledges support
from the ERC TOPONANOP (726001), the government of Spain (PID2019-106875GB-I00)
and Generalitat de Catalunya (CERCA, AGAUR, 2021 SGR 01443). Furthermore, the
research leading to these results has received funding from the European Union’s
Horizon 2020 under grant agreement no. 881603 (Graphene flagship Core3) and
820378 (Quantum flagship).

\bibliography{references,manual-references}

\end{document}